\documentclass[3p,times]{elsarticle}
 
\biboptions{comma,square}

\usepackage[figuresright]{rotating}

\newcommand{\pt}{ p_{\rm t}}

\begin{document}

\begin{frontmatter}

\title{ A Theoretic Review of Centrality-Dependent Direct Photon $\pt$ spectra in Au+Au Collisions at $\sqrt{s_{NN}}$=200 GeV  }

\author {Fu-Ming Liu}  

\address {Institute of Particle Physics, Central China Normal University, Wuhan, China}

\begin{abstract}
The recently published PHENIX data of direct photons at low $\pt$ from Au+Au collisions at the BNL relativestic Heavy Ion Collider (RHIC) energy 
 $\sqrt{s_{NN}}$=200 GeV  have attracted a lot of theoretic interest. Here I review our theoretical calculation which successfully reproduces the measured $\pt$ spectra of direct photons from Au+Au at different centralities over a wide $\pt$ range, in order to show a close relation between hadronic observables and those of direct photons.
\end{abstract}


\end{frontmatter}

\section{Introduction}
\label{} 
Recently PHENIX collaboration\cite{PHENIX08g} published a very precise direct photon pt spectra from  Au+Au collisions at  (RHIC) energy  $\sqrt{s_{NN}}$=200 GeV with $\pt$ interval from 1 to 5 GeV/c, based on an interpolation of virtual photons.
This  $\pt$ interval is of special importance at RHIC top energy, because thermal photons emitted from a plasma formed in heavy ion collisions play a dominant role.

At RHIC we usually separate the $\pt$ axis into three intervals:

1) high $\pt$ region: hadrons are mainly produced from the fragmentation of hard partons, at the early stage of the collision system. At the same stage, prompt photons are the main source of direct photons.
Different from hadrons, photons can couple to quarks through QED vertex.
So prompt photons can be either leading order photons from initial parton scattering, or fragmentation photons similar to fragmentation hadrons.

2) low $\pt$ region:  particles (hadrons or direct photons) are mainly emitted from thermal bath in hydrodynamical models (macroscopic model), or from secondary scatterings in cascade models (microscopic models), at a late stage of the collision system.
  
3) intermediate $\pt$ region: the scattering between hard partons and  thermal partons play an important role here. So the recombination of thermal partons and hard partons contributes a lot to hadron production, while jet-photon conversion(JPC) is an important source of direct photons.

At SPS energy, ie, Pb+Pb at 158GeV of lab frame, prompt photons and thermal photons are overlapped below 5~GeV/c with close slopes, so the three $\pt$ intervals can not be separated.
At RHIC the intermediate $\pt$ region is about [4,6]GeV/c.
At LHC, the measured highest $\pt$ can reach 150GeV/c, and the bulk physics still remains below 4 GeV/c. So a very wide intermediate $\pt$ region may appear.

A lot of theoretic calculation has been done to in order to understand the direct photon $\pt$ spectra measured by PHENIX or in order to obtain the plasma information from those data\cite{Bauchle:2010ym}. 
Among them, our recent theoretical calculation\cite{FML} could
reproduce the data relatively better and over a wider $pt$ region.
Here we review the calculation, in order to understand the close relation between hadronic observables and direct photons'. In the following section 2, the calculation approach will be reviewed. Main results will be reviewed in section 3. Conclusions and discussion will be made in section 4.

\section{Approach}
\label{} 

In order to calculate direct photon production from heavy ion collision over a wide pt range, we have to pay attention to two things at each collision stage, the reaction dynamics and the structure of the collision system.

At partonic level, the reactions relevant to direct photon production are the following processes: $qq\bar \rightarrow g \gamma$, $q g \rightarrow q \gamma$ and $q \rightarrow q \gamma$ where a quark can be replaced by an antiquark; in a hadronic gas,  direct photons can also be produced from the reactions such as $\pi \pi \rightarrow \rho \gamma $, $\pi\rho  \rightarrow \pi  \gamma $  and so on.

In our calculation, two stages were focused,  
the early stage when the two accelerated gold nucleus are overlapped and  the late stage  where the collision system is assumed in a local thermal equilibrium. The temperature and the flow velocity (related to the lab frame) of the thermal source at each space point evolutes with time, which is  governed by conservation laws of energy and momentum\cite{Hirano}
$\partial_{\mu}T^{\mu\nu}=0$
with the energy-momentum tensor 
$T^{\mu\nu}=(e+P)u^{\mu}u^{\nu}-Pg^{\mu\nu}$
for the perfect fluid. 
The pre-evolution of the system, usually studied with parton cascade,  and post-evolution stage of the dilute system, usually studied with hadron cascade, were not included in our work. 

At the early stage prompt photons are produced, where the partons at the left hand of the reaction dynamics are obtained from nuclear parton distribution function based on a lot of experimental data and theoretical study. 
 
At the late stage, the partons at the left hand of the reaction dynamics are obtained from a thermal (statistical) distribution at given temperature. The produced photons, first at the local rest frame (where the thermal source at rest), should be boost to the lab frame with the flow velocity.

The following sources of direct photons in heavy ion collisions were considered in our calculation:

1) Prompt photons, which includes the leading order photons(LO)
 \begin{eqnarray}
  \frac{dN^{\rm LO}}{dyd^{2}\pt}=T_{AB}(b)\sum_{{\displaystyle ab}}\int dx_{a}dx_{b}G_{a/A}(x_{a},M^{2})  G_{b/B}(x_{b},M^{2})\frac{\hat{s}}{\pi}\frac{d\sigma}{d\hat{t}}(ab\rightarrow\gamma+X)\delta(\hat{s}+\hat{t}+\hat{u})\label{eq:ABtogamma}\end{eqnarray}
and fragmentation photons(Frag)
 \begin{equation}
\frac{dN^{\mathrm{Frag}}}{dyd^{2}\pt}=
KT_{AB}(b)\sum_{ab,c=g,q_{i}}\int dx_{a}dx_{b}G_{a/A}(x_{a},M^{2}) 
 G_{b/B}(x_{b},M^{2})\frac{\hat{s}}{\pi}\frac{d\sigma}{d\hat{t}}
(ab\rightarrow cd)\delta(\hat{s}+\hat{t}+\hat{u})
\int dz_{c}\frac{1}{z_{c}^{2}}D_{\gamma/c}(z_{c},Q^{2}). 
\label{eq:jet-gamma0}\end{equation}
where $K=2$ for high order contributions to hard parton production and
$T_{AB}(b)$ is nuclear overlapping function. Energy loss of each hard parton was considered in fragmentation function $D_{\gamma/c}(z_{c},Q^{2})$,  according to its trajectory in the expanding plasma and the scattering with thermal partons. The parameter of energy loss per unit distance was determined so that pion  nuclear suppression factor $R_{AA}^{\pi}$ data at high $\pt$ were reproduced at all centralities.

2) Thermal photon production was calculated as a space-time integral of 
 emission rate:
\begin{equation}
\frac{dN^{\rm thermal}}{dy\, d^{2}p_{t}}=\int d^{4}x\,\Gamma(E^{*},T).
\label{eq1}
\end{equation}
where   $\Gamma(E^{*},T)$  is the thermal photon emission rate, which covers the contributions from both QGP phase \cite{Kapusta1991,AMY2001}
and HG phase \cite{Kapusta1991,Rapp2004},   $d^{4}x$  is the space-time volume unit and
$E^{*}=p^{\mu}u_{\mu}$ with $p^{\mu}$ and $u^{\mu}$ are 4-momentum of emitted photon and the 4-velocity of thermal source in the laboratory frame.  According to transport theory, photons emission rate in QGP(hadronic gas) is a convention of thermal parton(hadron) distribution and the reaction dynamics relevant to direct photon production at parton(hadron) level mentioned above.

3) JPC photons come from the scattering between hard partons and thermal partons when the hard partons penetrate in the thermal source. Our calculation was done in a similar way of thermal photons,
an integration of emission rate over the
space-time evolution of the hot and dense matter in the QGP phase
\begin{equation}
\frac{dN^{{\rm jpc}}}{dyd^{2}\pt}=\int\Gamma^{\rm jpc}(E^{*},T)f_{{\rm QGP}}(x,y,\eta,\tau)d^{4}x,\label{eq:jpc_int}\end{equation}
where the photon emission rate\cite{WangCY,Fries2005} $\Gamma^{\rm jpc}(E^{*},T)$ is a convention of a thermal parton distribution, a hard parton distribution, and the reaction dynamics relevant to direct photon production at parton level, and $f_{\rm QGP}$ is the fraction of QGP phase in the thermal bath. 
Energy loss of each hard parton was again considered according to its trajectory in the expanding plasma, similar to fragmentation contribution, which modified hard parton phase space distribution and 
 the JPC photon emission rate. 

In proton proton(pp) collisions at RHIC energy, there is no QGP formation. Therefore only the first source, the prompt photons contribute
to direct photon production. No hard parton energy loss there. 
So the correspond $\pt$ spectrum is read
 \begin{eqnarray}
    \frac{dN^{\rm pp}}{dyd^{2}\pt} & = & \frac{dN^{\rm LO}}{dyd^{2}\pt} + \frac{dN^{\rm Frag}}{dyd^{2}\pt} \\ \nonumber
 & = & \frac{1}{\sigma_{\rm inel}^{\rm pp}} \sum_{{\displaystyle ab}} \int dx_{a}dx_{b}G_{a/p}(x_{a},M^{2})  G_{b/p}(x_{b},M^{2})\frac{\hat{s}}{\pi}\frac{d\sigma}{d\hat{t}}(ab\rightarrow\gamma+X)\delta(\hat{s}+\hat{t}+\hat{u})  \\
 & + &  \frac{1}{\sigma_{\rm inel}^{\rm pp}} K  \sum_{ab,c=g,q_{i}}  \int dx_{a}dx_{b}G_{a/A}(x_{a},M^{2}) 
 G_{b/B}(x_{b},M^{2})\frac{\hat{s}}{\pi}\frac{d\sigma}{d\hat{t}}
(ab\rightarrow cd)\delta(\hat{s}+\hat{t}+\hat{u})
\int dz_{c}\frac{1}{z_{c}^{2}}D_{\gamma/c}(z_{c},Q^{2}). \nonumber
\label{eq:pp-gamma0} \end{eqnarray}

Nuclear modification factor of direct photons is calculated as 
$R_{AB}^{\gamma} = \frac{dN^{\rm Total}_{\rm AB}}{dy\, d^{2}p_{t}} /   \frac{dN^{\rm pp}}{dyd^{2}\pt} / N_{\rm coll}(b). $

\section{Results}
\label{} 
In the following our main results concerning to $\pt$ spectra are collected.

\begin{figure}
\includegraphics[scale=0.7]{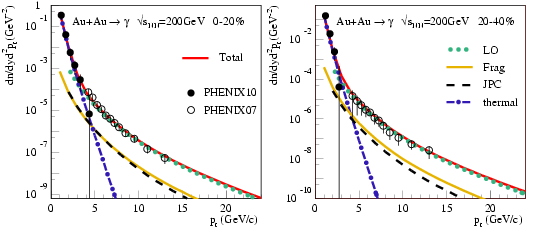}
\caption{\label{fig:c12-34} (Color Online) 
Direct photon production in Au+Au collisions at centrality 0-20\% and 20-40\%. 
PHENIX data are shown as open circles~\cite{PHENIX07PRL } and filled circles~\cite{PHENIX08g}.
Contributions of different sources (LO, Frag, JPC and thermal), are plotted with different line types, and added to get the total contribution(Total), which coincides quite well with experimental data points.
}
\end{figure}

1) Direct photon pt spectra in Au+Au collisions at centrality 0-20\% and 20-40\%.  
are presented in Fig.~\ref{fig:c12-34}.  PHENIX data are shown as open circles~\cite{PHENIX07PRL } and filled circles~\cite{PHENIX08g}.
Contributions of different sources (LO, Frag, JPC and thermal), are plotted with different line types, and added to get the total contribution(Total), which coincides quite well with experimental data points.

\begin{figure*}
\includegraphics[scale=0.7]{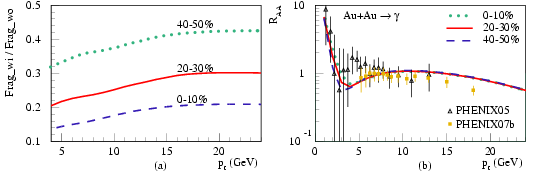}

\caption{\label{fig:Eloss} (Color Online)(a)The ratio of the fragmentation contribution with energy loss to the one without.
(b)Direct photon nuclear modification factor at three centralities, 
0-10\%, 
20-30\% and 40-50\% are illustrated as different line types. Data for 0-10\% centrality
are from PHENIX\cite{PHENIX data}.}
\end{figure*}

2) Hard parton energy loss in the plasma suppresses the fragmentation contributions and jet-photon conversion. This effect has been included in the curve of Frag and JPC contribution of Fig.1. 
In fact this effect can be illustrated via the ratios of 
the contribution with energy loss to the one without, 
as shown in Fig.~\ref{fig:Eloss}(a). Energy
loss in the plasma depends on the path length of the hard parton inside
the plasma, which turns out to depend on the collision centrality.
We do see a similar centrality dependence to charged hadron suppression 
 (jet quenching effect) in fragmentation contribution. However,energy loss effect does not show up in total $\pt$ spectra. Instead, we got an centrality-independent $R_{AA}^{\gamma}$, as shown in  Fig.~\ref{fig:Eloss}(b). A strong enhancement of $R_{AA}^{\gamma}$ appears  at low $\pt$ due the thermal contribution.

\section{Conclusion}
\label{} 
In this paper I shortly review our calculation of direct photon production at RHIC energy.
We reproduced PHENIX low $\pt$ direct photon data successful, due to two factors:
One is the structure of the collision system, a hot dense plasma created in a heavy ion collision which is the thermal source of photon emission. This system is described with ideal hydrodynamic equation with relevant parameters constrained by hadronic experimental data. The other factor is reaction dynamics of thermal photon production. Here we take AMY photon emission rate instead of Kapusta's photon emission rate. Due to LPM effect, AMY photon emission rate is higher than Kapusta's and let the plasma shine stronger. 

At high $\pt$ region, both the initial effect (isospin in nuclear parton distribution factor) and the final effect (hard parton energy loss in plasma) work. However, the main difference between direct photon production and hadron production is that, additional to the fragmentation contribution, there is a leading order contribution to direct photons. And at high $\pt$, perturbative theory works better and makes leading order contribution absolutely dominant. This term does not effected by energy loss, so that the nuclear modification factor of direct photons is at the magnitude of unity. It goes down slowly due to isospin effect.

At low $\pt$, a strong enhancement of $R_{AA}^{\gamma}$ appears due the thermal contribution. So a strong enhancement of direct photons at low $\pt$  compared pQCD calculation of prompt photons can be a QGP formation signal.   

When the $\pt$ spectra of direct photons at different centralities were successfully reproduced, has everything concerning to direct photons been understood? No. We got only the mean production at each given $\pt$ value. Other observables, such as elliptic flow, correlation, and so on, will provide additional information than $\pt$ spectra.

\section*{Acknowledgments}
This work is supported by the Natural Science Foundation of China
under the project No.~10975059.

\end{document}